\documentclass[prl,shwopacs,amsmath,twocolumn]{revtex4}
\usepackage{amssymb}
\usepackage{graphics}

\begin{document}

\newcommand{\one}{\ensuremath{\hbox{$\mit I$\kern-.3em$\mit I$}}}
\renewcommand{\one}{\ensuremath{\hbox{$\mathrm I$\kern-.6em$\mit 1$}}}
\renewcommand{\one}{\ensuremath{\hbox{$\mathrm I$\kern-.6em$\mathrm 1$}}}
\newcommand{\ad}{\ensuremath{a^\dagger}}
\newcommand{\bd}{\ensuremath{b^\dagger}}
\newcommand{\cd}{\ensuremath{c^\dagger}}
\newcommand{\dd}{\ensuremath{d^\dagger}}
\newcommand{\ed}{\ensuremath{e^\dagger}}
\newcommand{\ket}[1]{\ensuremath{|{#1}\rangle}}
\newcommand{\bra}[1]{\ensuremath{\langle{#1}|}}
\newcommand{\braket}[1]{\ensuremath{\langle{#1}\rangle}}

\title{Quantum computation with unknown parameters}

\author{J.~J. \surname{Garc\'{\i}a-Ripoll}}
\affiliation{Max-Planck-Institut f\"ur Quantenoptik,
  Hans-Kopfermann-Str. 1, Garching, D-85748, Germany.}
\author{J.~I. \surname{Cirac}}
\affiliation{Max-Planck-Institut f\"ur Quantenoptik,
  Hans-Kopfermann-Str. 1, Garching, D-85748, Germany.}

\date{\today}

\begin{abstract}
  We show how it is possible to realize quantum computations on a system in
  which most of the parameters are practically unknown.  We illustrate our
  results with a novel implementation of a quantum computer by means of bosonic
  atoms in an optical lattice. In particular we show how a universal set of
  gates can be carried out even if the number of atoms per site is uncertain.
\end{abstract}

\maketitle

Scalable quantum computation requires the implementation of quantum gates with
a very high fidelity. This implies that the parameters describing the physical
system on which the gates are performed have to be controlled with a very high
precision, something which it is very hard to achieve in practice. In fact, in
several systems only very few parameters can be very well controlled, whereas
other posses larger uncertainties. These uncertainties may prevent current
experiments from reaching the threshold of fault tolerant quantum computation
(i.e. gate fidelities of the order of $0.9999$ \cite{tolerancia}), that is, the
possibility of building a scalable quantum computer. For example, in quantum
computers based on trapped ions or neutral atoms \cite{cold-atoms}, the
relative phase of the lasers driving a Raman transition can be controlled very
precisely, whereas the corresponding Rabi frequency $\Omega$ has a larger
uncertainty $\Delta\Omega$.  If we denote by $T$ the time required to execute a
gate (of the order of $\Omega^{-1}$), then a high gate fidelity requires
$T\Delta\Omega\ll 1$ (equivalently, $\Delta\Omega/\Omega\ll 1$), which
may very hard to achieve, at least to reach the above mentioned threshold.

In this letter we show how to achieve a very high gate fidelity even when most
of the parameters describing the system cannot be adjusted to precise values.
Our method is based on the technique of adiabatic passage, combined with some
of the ideas of quantum control theory.  We will illustrate our method with a
novel implementation of quantum computing using atoms confined in optical
lattices \cite{QC-mott}. If the number of atoms in each of the potential well
is uncertain, which is one of the problems with this kind of experiments,
most of the parameters will have an uncertainty of the order of the parameter
itself (e.g. $|\Delta\Omega|\sim \Omega$), which under normal circumstances
will give rise to very poor fidelities and even impede the performance of
quantum gates. As we will show, using our method not only quantum computation
is possible but even very high fidelities could be achieved.

The use of adiabatic passage techniques to implement quantum gates is, of
course, not a new idea. In fact, several methods to perform certain quantum
gates using Berry phases have been put forward recently
\cite{holonomias2,holonomias,Duan}.  Furthermore, adiabatic passage techniques
have been proposed as a way of implementing a universal set of holonomies
\cite{Duan}, i.e. quantum gates which are carried out by varying certain
parameters and whose outcome only depends on geometrical properties of the
paths in parameter space \cite{holonomias2}. In all these proposals, physical
implementations of standard quantum computation have been adapted so that the
quantum gates are performed in an adiabatic way giving rise to holonomies.
Despite its clear fundamental interest, it is not clear yet if such a novel way
of implementing the quantum gates may offer real benefits with respect to the
original proposals. In contrast, in our illustrative example adiabatic passage
is required to perform quantum gates and therefore it is an essential tool not
only to achieve the desired precision but also to build a quantum computer at
all.

The outline of the paper is as follows. First we will show how to produce a
universal set of gates (Hadamard, phase, and CNOT) starting from two
Hamiltonians in which only one parameter is precisely controlled. Then we will
show that this method eliminates an important obstacle in a particular physical
scenario that has been proposed for quantum computation \cite{QC-mott}, namely
a set of atoms in optical lattices interacting via cold collisions.

Let us consider a set of qubits that can be manipulated according to the single
qubit Hamiltonian
\begin{equation}
\label{H1} H_1 = \frac{\Delta}{2} \sigma_z +
\frac{\Omega}{2}(\sigma_+ e^{i\varphi} + \sigma_- e^{-i\varphi}),
\end{equation}
and the two--qubit Hamiltonian
\begin{equation}
  \label{H2}
  H_2 = \tilde\Delta \ket{11}\bra{11} +
  \frac{\tilde \Omega}{2}\; \one \otimes
      (\sigma_+ e^{i\varphi} + \sigma_- e^{-i\varphi}).
\end{equation}
As mentioned above, we will assume that most of the parameters appearing in
these Hamiltonians are basically unknown. On the other hand, we will not
consider any randomness in these parameters (the corresponding errors could be
corrected with standard error correction methods \cite{Nielsen}, as long as
they are small).  In particular we will assume that only $\varphi$ can be
precisely controlled.  For the other parameters we will assume that: (1) they
are given by an unknown (single valued) function of some experimentally
controllable parameters; (2) they can be set to zero. For example, we may have
$\Omega=f(I)$, where $I$ is a parameter that can be experimentally controlled,
and we only know about $f$ that $f(0)=0$ and that we can reach some value
$\Omega_m\equiv f(I_m) \neq 0$ for some $I_m$ \cite{unknown}.  Below we analyze
a particular physical scenario which exactly corresponds to this situation.
However, we want to stress that this situation can be naturally found in more
general scenarios.  For example, for atomic qubits the states $|0\rangle$ and
$|1\rangle$ may correspond to two degenerate atomic (ground state) levels which
are driven by two lasers of the same frequency and different polarization. The
corresponding Hamiltonian is given by (\ref{H1}), where the parameters
$\varphi,\Omega,\Delta$ describe the relative phase of the lasers, the Rabi
frequency and detuning of the two-photon Raman transition, respectively. The
Rabi frequency can be changed by adjusting the intensity of the lasers, and the
detuning and the phase difference by using appropriate modulators.  In
practice, $\Omega$ ($\Delta$) can be set to zero very precisely by switching
off the lasers (modulators) and $\varphi$ can be very precisely controlled to
any number between 0 and $2\pi$. However, fixing $\Omega$ or $\Delta$ to a
precise value (for example, $23.098$ kHz) may be much more challenging.

\begin{figure}[t]
  \centering \resizebox{\linewidth}{!}{\includegraphics{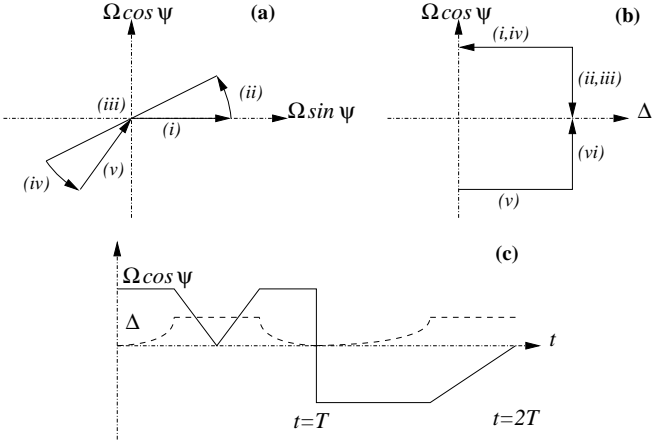}}
  \caption{
    \label{fig-local}
    Schema of how the parameters of Hamiltonian (\ref{H1}) have to be changed
    in order to perform a phase gate (a), and Hadamard gate (b). In Fig. (c)
    we show a possible evolution of the parameters $\Delta$ and $\Omega$
    for the Hadamard gate.}
\end{figure}

The idea of obtaining perfect gates with unknown parameters relies on the
combination of adiabatic passage techniques \cite{holonomias2} and ideas of
quantum control \cite{Bang}. Let us recall the basic idea of adiabatic passage.
Suppose we have a Hamiltonian that depends parametrically on a set of
parameters, denoted by $p$, which are changed adiabatically with time along a
given trajectory $p(t)$. After a time $T$, the unitary operator corresponding
to the evolution is given by
\begin{equation}
U(T)= \sum_\alpha e^{i(\phi_\alpha + \psi_\alpha)}
|\Phi_\alpha[p(T)]\rangle\langle \Phi_\alpha[p(0)]|.
\end{equation}
Here, $|\Phi_\alpha(p)\rangle$ are the eigenstates of the Hamiltonian for which
the parameters take on the values $p$. The phase $\phi_\alpha$ is a dynamical
phase that explicitly depends on how the parameters $p$ are changed with time,
whereas the phase $\psi_\alpha$ is a purely geometrical phase ands depends on
the trajectory described in the parameter space.  Our basic idea to perform any
given gate is first to design the change of the parameters in the Hamiltonians
(\ref{H1})-(\ref{H2}) such that the eigenvectors evolve according to the
desired gate, and then to repeat the procedure changing the parameters
appropriately in order to cancel the geometric and dynamical phases.

Let us first show how to perform the phase gate $U=e^{i\theta\sigma_z/2}$. We
set $\Delta=0$ for all times. The parameters $(\Omega,\varphi)$ have to be
changed as follows [see Fig.\ \ref{fig-local}(a)] :
\begin{eqnarray}
(0,0) &\stackrel{(i)}{\rightarrow}& (\Omega_m,0)
\stackrel{(ii)}{\rightarrow} (\Omega_m,\theta/2)
\stackrel{(iii)}{\Rightarrow} (\Omega_m,\theta/2+\pi) \nonumber \\
&\stackrel{(iv)}{\rightarrow}& (\Omega_m,\theta+\pi)
\stackrel{(v)}{\rightarrow} (0,\theta+\pi)
\end{eqnarray}
Steps (i,ii) and (iv,v) are performed adiabatically and require a time $T$.
The double arrow of step (iii) indicates a sudden change of parameters.  Note
that $\Omega(0)=\Omega(2T)=0$, $\Omega(t)=\Omega(2T-t)$ and $\varphi(t) = \pi +
\theta - \varphi(2T-t)$, which does not require the knowledge of the function
$f$ but implies a precise control of the phase.  A simple analysis shows that
(i-v) achieve the desired transformation $\ket{0} \to e^{i\theta/2} \ket{0}$,
$\ket{1} \to e^{-i\theta/2} \ket{1}$.  Note also that the dynamical and
geometrical phases acquired in the adiabatic processes (i-v) cancel out.

The Hadamard gate can be performed in a similar fashion. In the space of
$[\Delta,\Omega_x=\Omega\cos(\varphi)]$, the protocol is
\begin{eqnarray}
  (0,\Omega_m) &\stackrel{(i)}{\rightarrow}&
  (\Delta_m,\Omega_m) \stackrel{(ii)}{\rightarrow}
  (\Delta_m,0) \stackrel{(iii)}{\rightarrow}
  (\Delta_m,\Omega_m) \stackrel{(iv)}{\rightarrow} \nonumber\\
  (0,\Omega_m) &\stackrel{(v)}{\Rightarrow}&
  (0,-\Omega_m) \stackrel{(vi)}{\rightarrow}
  (\Delta,-\Omega_m) \stackrel{(vii)}{\rightarrow}
  (\Delta,0),
\end{eqnarray}
as shown in Fig.\ \ref{fig-local}(b-c).  In order to avoid the dynamical
phases, we have to make sure that steps (i-v) are run in half the time as
(vi-vii).  More precisely, if $t<T$, we must ensure that
$\Delta(t)=\Delta(T-t)$, $\Omega_x(t)=\Omega_x(T-t)$, $\Delta(T+t) =
\Delta(t/2)$ and $\Omega_x(T+t) = \Omega_x(t/2)$. With this requisite we get
$\frac{1}{\sqrt{2}}(\ket{0} + \ket{1}) \to \ket{0}$,
$\frac{1}{\sqrt{2}}(\ket{0} - \ket{1}) \to -\ket{1}$.  Again, the whole
procedure does not require us to know $\Omega$ or $\Delta$, but rather to
control the evolution of the experimental parameters which determine them.

The C-NOT gate requires the combination of two two-qubit processes using $H_2$
and one local gate. The first process involves changing the parameters
$[\tilde\Delta,\tilde\Omega_x=\tilde\Omega\cos(\varphi)]$ of Eq. (\ref{H2})
according to
\begin{eqnarray}
(\tilde\Delta_m,0) &\stackrel{(i)}{\rightarrow}&
(\tilde\Delta_m,\tilde\Omega_m)
\stackrel{(ii)}{\rightarrow} (0,\tilde\Omega_m)
\stackrel{(iii)}{\Rightarrow} (0,-\tilde\Omega_m) \nonumber \\
&\stackrel{(iv)}{\rightarrow}& (\tilde\Delta_m,-\tilde\Omega_m)
\stackrel{(v)}{\rightarrow} (\tilde\Delta_m,0). \label{sigmay}
\end{eqnarray}
This procedure gives rise to the transformation
\begin{equation}
  U_1 = \ket{0}\bra{0}\otimes \one + e^{i\xi}\ket{1}\bra{1}\otimes i\sigma_y,
\end{equation}
where $\xi = \int_0^T \delta(t) dt$ is an unknown dynamical phase. The second
operation required is a NOT on the first qubit $U_2 = (\ket{0}\bra{1} +
\ket{1}\bra{0}) \otimes \one$.  Finally, if $\tilde\Delta^{(1)}(t)$ denotes the
evolution of $\tilde\Delta$ in Eq.  (\ref{sigmay}), we need to follow a path
such that $\tilde\Delta^{(3)}(t) = \tilde\Delta^{(1)}(t)$,
$\tilde\Omega^{(3)}(t) = 0$.  If the timing is correct, we achieve the
transformation
\begin{equation}
  U_3 = (\ket{0}\bra{0} + e^{i\xi}\ket{1}\bra{1})\otimes \one.
\end{equation}
Everything combined gives us the CNOT up to a global unimportant phase
$U_{cnot} = \ket{0}\bra{0}\otimes \one +\ket{1}\bra{1}\otimes i\sigma_y
  = e^{-i\xi}U_2 U_3 U_2 U_1$.

\begin{figure}[t]
  \centering
  \resizebox{\linewidth}{!}{\includegraphics{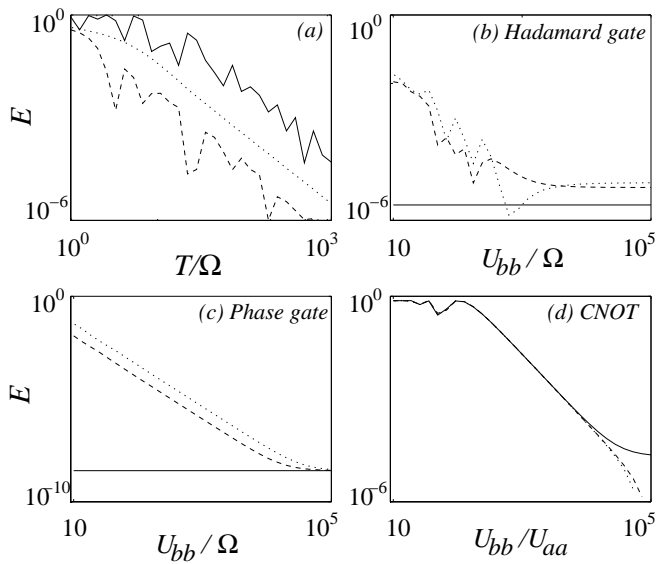}}
  \caption{\label{fig-error}
    Log-log plot of the gate error, $E=1-{\cal F}$, for different local and
    nonlocal gates. (a) Realization based on Hamiltonians
    (\ref{H1})-(\ref{H2}), with fixed parameters and varying time. We plot the
    error for the Hadamard (solid), phase (dashed) and CNOT gates (dotted).
    (b-d) Errors for our proposal with atomic ensembles. We fix the the total
    time all parameters except the interaction constant $U_{bb}$. In (b-c) we
    plot realizations for $n=1$ (solid) and $n=3,5$ (dashed) atoms per lattice
    site. In (d) we plot simulations with a difference of atoms of
    $|n-m|=0,1,2$ between both wells (solid, dash and dotted lines). The vaues
    of the parameters are given in the text.}
\end{figure}

In Fig.\ \ref{fig-error}(a) we illustrate the performance of our method, as
well as its sensitivity against non--adiabatic processes. As a figure of merit
we have chosen the gate fidelity \cite{Nielsen} ${\cal F} =
|\mathrm{Tr}\{U_{ideal}^\dagger U_{real}\}|^2/d^2$, where $d$ is the
dimensionality of the space ($2$ for local gates, $4$ for two-qubit gates),
$U_{ideal}$ is the gate that we wish to produce and $U_{real}$ is the actual
operation performed.  As expected, for fixed parameters $\{\Delta_m/\Omega_m =
\tilde\Delta_m/\tilde\Omega_m = 1/10, \varphi_m=\pi/4\}$, the adiabatic theorem
applies when the processes are performed with a sufficiently slow speed.
Typically a time $T\sim 300/\Omega_m, 300/\tilde\Omega_m$ is required for the
desired fidelity ${\cal F}=1-10^{-4}$.


Let us now consider a set of bosonic atoms confined in a periodic optical
lattice at sufficiently low temperature (such that only the first Bloch band is
occupied). The atoms have two relevant internal (ground) levels, $|a\rangle$
and $|b\rangle$, in which the qubit is stored. This set-up has been considered
in Ref.  \cite{QC-mott} where it has been shown how single quantum gates can be
realized using lasers and two--qubit gates by displacing the atoms that are in
a particular internal state to the next neighbor location.  The basic
ingredients of such a proposal have been recently realized experimentally
\cite{Bloch2}. However, in this and all other schemes so far \cite{williams} it
is assumed that there is a single atom per lattice site since otherwise even
the concept of qubit is no longer valid. In present experiments, in which the
optical lattice is loaded with a Bose-Einstein condensate
\cite{Jaksch1,Bloch1}, this is not the case (since zero temperature is required
and the number of atoms must be identical to the number of lattice sites). We
show now a novel implementation in which, with the help of the methods
presented above, one overcomes this problem.

For us \textit{a qubit will be formed by an aggregate of atoms at some lattice
  site}. The number of atoms forming each qubit is completely unknown. The only
requirement is that there is at least one atom per site \cite{superfluid}. We
will denote by $n_k$ the number of atoms in the $k$--th well and identify the
states of the corresponding qubit as
\begin{equation}
\ket{0}_k = \frac{a_k^{\dagger n_k}}{\sqrt{n!}} \ket{{\text vac}},\quad
\ket{1}_k = \bd_k\frac{a_k^{\dagger (n_k-1)}}{\sqrt{(n_k-1)!}}
\ket{{\text vac}},
\end{equation}
where $a_k^\dagger$ ($b_k^\dagger$) are the creation operators for one atom in
levels $|a\rangle$ and $|b\rangle$, respectively. The quantum gates will be
realized using lasers, switching the tunneling between neighboring sites, and
using the atom--atom interaction.

In the absence of any external field, the Hamiltonian describing our system is
\begin{equation}
\label{H-orig}
 H =- J^{(b)}_{k} \sum_k (\bd_{k+1}b_k + \bd_kb_{k+1})
 + \frac{U_{bb}}{2} \sum_k \bd_k\bd_kb_kb_k.
\end{equation}
Here, $U_{bb}$ and $J^{(b)}_k$ describe the interactions between and the
tunneling of atoms in state $|b\rangle$. We will assume that $J^{(b)}_k$ can be
set to zero and increased by adjusting the intensities of the lasers which
create the optical lattice. We have assumed that the atoms in state $|a\rangle$
do not interact at all and do not hop, something which may be achieved by
tuning the scattering lengths and the optical lattice. Both restrictions will
be relaxed later on. The Hamiltonian (\ref{H-orig}) possesses a very important
property when all $J^{(b)}_k=0$, namely it has no effect on the computational
basis (i.e. $H|\Psi\rangle=0$ for all states $|\Psi\rangle$ in the Hilbert
space generated by the qubits).  Otherwise, it would produce a non--trivial
evolution that would spoil the computation.

We show now how a single qubit gate on qubit $k$ can be realized using lasers.
First, during the whole operation we set $J^{(b)}_k=0$ in order to avoid
hopping. The laser interaction is described by the Hamiltonian
\begin{equation}
  \label{H1-real}
  H_{las}^{(k)} = \frac{\Delta_k}{2} (\ad_k a_k - \bd_k b_k) +
  \frac{\Omega_k}{2} (e^{i\varphi}\ad_k b_k + e^{-i\varphi}\bd_k a_k).
\end{equation}
For $U_{bb}\gg |\Delta_k|,|\Omega_k|$, we can project the total Hamiltonian
acting on site $k$ by an effective Hamiltonian acting on the qubit which
resembles (\ref{H1}), with $\Delta=\Delta_k$ and $\Omega =
\Omega_k\sqrt{n_k-1}$.  Thus, using the methods exposed above we can achieve
the Hadamard and phase gates with a high precision, even though the coupling
between the bosonic ensemble and light depends on the number of atoms.

For the realization of the two-qubit Hamiltonian (\ref{H2}) we need to combine
several elements. First of all we need the Raman coupling of Eq.
(\ref{H1-real}) to operate on the second well. Second, we need to tilt the
lattice using a magnetic field \cite{Bloch1} which couples to states $\ket{a}$
and $\ket{b}$ differently, $H_{tilt} = \sum_k k g(\ad_ka_k - \bd_kb_k)$.  And
finally we must allow virtual hopping of atoms of the type $\ket{b}$ ($J^{(b)}
\ll |U_{bb}-g|$). After adiabatic elimination we find that the effective
Hamiltonian depends on the number of particles in the second site, $n_2$,
\begin{equation}
  \label{H2-eff}
  H_2^{eff}= \frac{J_b^2}{g-U_{bb}} \ket{11}\bra{11} +
  \sqrt{n_2-1}\tilde\Omega\;\one\otimes \sigma_x.
\end{equation}
The identification with Eq. (\ref{H2}) is evident, and once more the use of
adiabatic passage will produce gates which are independent of the number of
particles.

We have studied the different sources of error which may affect our proposal:
(i) $U_{bb}$ is finite and (ii) atoms in state $\ket{a}$ may hop and interact.
These last phenomena are described by additional contributions to Eq.
(\ref{H-orig}) which are of the form $J^{(a)}_k(\ad_ka_{k+1}+\ad_{k+1}a_k)$,
$U_{aa}\ad_k\ad_ka_ka_k$, and $U_{ab}\ad_k\bd_ka_kb_k$. The consequences of
both imperfections are: (i) more than one atom per well can be excited, (ii)
the occupation numbers may change due to hopping of atoms and (iii) by means of
virtual transitions the effective Hamiltonian differs from (\ref{H1}) and
(\ref{H2}). The effects (i)-(ii) are eliminated if $(\Omega/U_{bb})^2\ll 1$ and
$(J^{(a)}_k/U_{bb})^2\ll 1$. Once these conditions are met, we may analyze the
remaining errors with a perturbative study of the Hamiltonians (\ref{H1-real})
and (\ref{H-orig}) plus the terms ($J^{(a)},U_{ab},U_{aa}$) that we did not
consider before. In Eq. (\ref{H1-real}), the virtual excitation of two atoms
increments the parameter $\Delta$ by an unknown amount, $\Delta_{eff} \sim
\Delta+ 2\Omega^2n_k / (\Delta+U_{ab}-U_{bb})$. If $U_{ab}\ll U_{bb}$ and
$\Omega^2n_k T/U_{bb} \ll 1$, this shift may be neglected.  In the two-qubit
gates the energy shifts are instead due to virtual hopping of all types of
atoms.  They are of the order of $\max(J^{(b)},J^{(a)})^2/g^2 \sim J^2/U_{bb}$,
and for $J^2T/U_{bb} \ll 1$ they also may be neglected.

To quantitatively determine the influence of these errors we have simulated the
evolution of two atomic ensembles with an effective Hamiltonian which results
of applying second order perturbation theory to Eq. (\ref{H-orig}), and which
takes into account all important processes. The results are shown in Figs.
\ref{fig-error}(c-d). For the two-qubit gate we have assumed $U_{aa}=U_{ab}$,
$J^{(a)}=J^{(b)}=J$, $J_{m}=0.05U_{ab}$, $\Omega_m=J_{m}^2/10$,
$g=U_{bb}+U_{ab}/2$ and operation time, $T=200/\Omega_{m}$, while changing the
ratio $U_{bb}/U_{ab}$ and the populations of the wells. For the local gates we
have assumed $\Delta_m=1$, $\Omega_m=U_{bb}/10$ and different occupation
numbers $n_k$, and we have also changed $U_{bb}/\Omega_m$.

We extract several conclusions. First, the stronger the interaction between
atoms in state $\ket{b}$, the smaller the energy shifts.  Typically, a ratio
$U_{bb} = 10^4 U_{ab}$ is required to make ${\cal F} = 1-10^{-4}$.  Second, the
larger the number of atoms per well, the poorer the fidelity of the local gates
[Figs.  \ref{fig-error}(b-c)]. And finally, the fidelity of the two-qubit gate
presents a small dependence on the population imbalance between wells.

In this work we have shown that it is possible to perform quantum computation
even when the constants in the governing Hamiltonians are unknown.  We have
developed a scheme based on performing adiabatic passage with one-qubit
(\ref{H1}) and two-qubit (\ref{H2}) Hamiltonians. With selected paths and
appropriate timing, it is possible to perform a universal set of gates
(Hadamard gate, phase gate and a CNOT). These procedures cannot only be used
for quantum computing but also for quantum simulation \cite{Guifre}. Finally,
based on the preceding ideas, we have proposed a scheme for quantum computing
with cold atoms in a tunable optical lattice. Our scheme works even when the
number of atoms per lattice site is uncertain. Note that, these ideas also
apply to some other setups like the microtraps recently realized in Ref.
\cite{Ertmer}.

We thank D. Liebfried and P. Zoller for discussions and the EU project EQUIP
(contract IST-1999-11053).


\begin{thebibliography}{99}
\bibitem{unknown}{Note that, in principle, one could measure the dependence of
    these parameters (function $f$). However in many realistic implementations
    this is not possible, since the measurements are destructive (lead to
    heating or the atoms escape the trap), and in different realizations the
    dependence is different.}
  
\bibitem{tolerancia}{A.~M. Steane, arXiv:quant-ph/0207119; P.~W. Shor, in {\em
      Proc. 35th Annual Symposium on Fundamentals of Computer Science} (IEEE
    Press, Los Alamitos, 1996), p.\ 56, arXiv:quant-ph/9605011; E. Knill, R.
    Laflamme, and W.~H. Zurek, Science {\bf 279}, 342 (1998); D.  Aharonov and
    M. Ben-Or, SIAM Jour. Comput. (submitted) arXive:quant-ph/9906129.}
  
\bibitem{cold-atoms}{J.~I. Cirac, and P. Zoller Phys. Rev. Lett. \textbf{74},
    4091 (1995); C. Monroe {\em et al.} Phys. Rev. Lett. \textbf{75}, 4714
    (1995).}

\bibitem{Bloch1}{M. Greiner {\em et al.}, Nature \textbf{415}, 39 (2002).}
  
\bibitem{Nielsen}{See, for example, M.~A. Nielsen and I.~L. Chuang, {\em
      Quantum Computation and Quantum Information} (Cambridge Univ. Press,
    Cambridge, 2002).}
  
\bibitem{Jaksch1}{D. Jaksch {\em et al.}, Phys. Rev. Lett. \textbf{81}, 3108
    (1998).}
  
\bibitem{QC-mott}{D. Jaksch {\em et al.}, Phys. Rev. Lett. \textbf{82}, 1975
    (1999).}
  
\bibitem{holonomias2}{P. Zanardi and M. Rasetti, Phys. Lett. A 264, 94 (1999);
    J. Pachos, P. Zanardi, M. Rasetti, Phys. Rev. A \textbf{61}, 010305(R)a
    (2000).}
  
\bibitem{holonomias}{J. A. Jones {\em et al}, Nature
    403, 869 (1999); G. Falci, et al., Nature 407, 355 (2000)}

\bibitem{Duan}{L.-M. Duan, J.~I. Cirac, and P. Zoller, Science \textbf{292},
    1695 (2001); A. Recati {\em et al.} , arXiv:quant-ph/0204030.}
  
\bibitem{Bang}{L. Viola, and S. Lloyd, Phys. Rev. A \textbf{58}, 2733 (1998);
    L.-M. Duan, and G. Guo, arXiv:quant-ph/9807072}

\bibitem{Bloch2}{I. Bloch (private communication).}

\bibitem{williams}{E. Charron {\em et al.}, Phys. Rev. Lett. 88, 077901 (2002);
    K. Eckert {\em et al.}, arXiv:quant-ph/0206096.}
  
\bibitem{superfluid}{The current scheme would even work with a superfluid phase
    which is abruptly loaded in a deep optical lattice.}

\bibitem{Guifre}{E. Jan\'e {\em et al.}, arXiv:quant-ph/0207011.}
  
\bibitem{Ertmer}{R. Dumke {\em et al.}, Phys. Rev. Lett. \textbf{89}, 097903
    (2002).}

\end{thebibliography}
\end{document}